# Description of chiral tetrahedral molecules *via* an Aufbau approach

Salvatore Capozziello[a,*] and Alessandra Lattanzi[b,*]

[a]*Dipartimento di Fisica "E. R. Caianiello", INFN sez. di Napoli, Università di Salerno, Via S. Allende 84081 Baronissi, Salerno, Italy*
[b]*Dipartimento di Chimica, Università di Salerno, Via S. Allende 84081 Baronissi, Salerno, Italy*

**Abstract** We propose a predictive "building-up" of tetrahedral molecules, based on a previously derived "chirality index" $\chi \equiv \{n, p\}$, which characterizes a tetrahedral molecule, with $n$ chiral centers, as achiral, diastereoisomer, or enantiomer as a function of the $p$ observed inversions (at most one for each center) under rotation and superimposition to its mirror immage. The set of rules which is identified allows the classification of the new molecule, obtained by the addition of another chiral center, by the determination of the selection rule $\Delta p = 0,1$ with respect to the added center.

1. **Introduction**

The development of mathematical descriptors for the characterization of chemical structures is a subject of increasing interest [1]. Several topological indexes have been proposed to depict 3D molecular structures and shapes [2]. Chirality of molecules has been as well the subject of studies aimed to achieve numerical indexes as a measurement of this property [3], so, discrete and continuous measurements of chirality have been proposed in order to determine the degree of chirality of a molecule. Such measurements are related to the methods for the characterization of physical and chemical features of a given compound.

---
[*]Corresponding authors. *E-mail:* lattanzi@unisa.it

On the other hand, the empirical classification of organic molecules is based on the properties of functional groups, such as hydroxy group in alcohols, C─C double and triple bonds, CO group in ketones, aldehydes, etc. In general, organic compounds are collected in homologous series, differing by the number of carbons present in the structure. The most important classification of organic molecules, from this point of view, is the Beilstein system, where each compound finds an indexed place in the *Beilstein Handbook of Organic Chemistry* [4]. Thermodynamic enthalpies of formation have been used as a basis of classification in homologous series [5]. Additivity schemes for atoms are introduced in order to predict the enthalpies for compounds of homologous series. Furthermore, quantum mechanical quantities such as molecular total and partial energies have been statistically treated for the same purpose [6].

In this paper, we want to propose a scheme, which allow to characterize tetrahedral chains as diastereoisomers, enantiomers or achiral molecules, by means of a building-up process (Aufbau-like). The approach is based on a geometrical representation of tetrahedral molecules by complex numbers [7] and on the $O(4)$ orthogonal group algebra [8] derived considering the well-known Fischer projections for tetrahedrons. The building-up process gives rise to a chirality index which assigns the intrinsic chiral structure of the final compound. The paper is organized as follow: in Section 2, a geometrical representation and the algebraic structure of tetrahedral chains are discussed. Section 3 is devoted to the description of chirality index, while the Aufbau rule is presented in Section 4. Discussion and conclusions are reported in Section 5.

## 2. Geometrical representation of tetrahedral chains and their algebraic properties

Chiral molecules, characterized by *central chirality*, have a given number of stereogenic centers. Two molecules with identical chemical formulas, if they are not superimposable, are called *enantiomers*. Two molecules with identical structural formulas and more than one chiral center, which are not mirror images of each other and not superimposable, are termed *diastereoisomers*. In the case of perfectly superimposable structures, we are dealing with achiral molecules.

In a recent report [7], we proposed a geometrical representation of a tetrahedral molecule starting from its projections on a given {x,y}-plane, where the molecule center coincides with the origin of axes.

The key ingredient of such an approach is the consideration that a given bond of a molecule can be represented by a complex number, where the "modulus" is the length of the bond with respect to the stereogenic center and the "anomaly" is the angular position with respect to the other bonds and the axes. Every bond, in the plane {x,y}, is

$$\Psi_j = \rho_j e^{i\theta_j} \qquad (1)$$

where $\rho_j$ is the projected length of the bond, $\theta_j$ is the angular position, having chosen a rotation versus; $i = \sqrt{-1}$ is the imaginary unit. A molecule with one stereogenic center (Figure 1) is then given by the sum vector

$$M = \sum_{j=1}^{4} \rho_j e^{i\theta_j} \qquad (2)$$

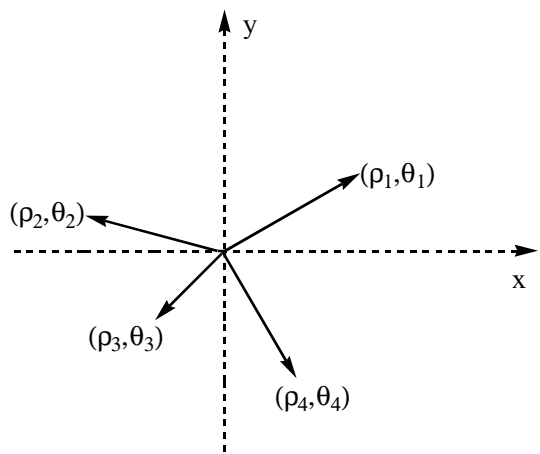

Fig. 1. Projection of a tetrahedral molecule on a plane containing the stereogenic center.

These considerations can be immediately extended to more general cases. A molecule with *n* stereogenic centers has *n* planes of projection and the bonds between two centers have to be taken into account. A molecule with one center has four bonds, a chain with *n* centers has $3n+1$ bonds. The rule works only for simply connected tetrahedrons so that a generic chain can be given by the relation

$$M_n = \sum_{k=1}^{n} \sum_{j=1}^{3n+1} \rho_{jk} e^{i\theta_{jk}} \tag{3}$$

where k is the "center-index", and j is the "bond-index". A projective plane of symmetry is fixed for any k and the couple $\{\rho_{jk}, \theta_{jk}\} \equiv \{0,0\}$ defines the center in every plane.

In another report [8], we worked out an algebraic approach in order to deal with central molecular chirality. Starting from the empirical Fischer rules, we developed an *O*(4) algebra of 24 matrix operators capable of characterizing all the representations of the two enantiomers of a given tetrahedral molecule. As a result, every chain of type (3) can be classified as an enantiomer, a diastereoisomer or an achiral molecule, thanks to the operators of such an algebra, which tell us if, with respect to a chosen fundamental representation of a given configuration for a tetrahedron, its bonds undergo a rotation or an inversion of two of their bonds.

Let us go into details to explain how such an algebra works. Let

$$M_k^{(0)} = \begin{pmatrix} \Psi_{1k} \\ \Psi_{2k} \\ \Psi_{3k} \\ \Psi_{4k} \end{pmatrix} \qquad (4)$$

be the column vector assigning the fundamental representation of the k-tetrahedron in a chain of $n$ tetrahedrons (3). The matrix operator

$$\chi_1^k = \begin{pmatrix} 1 & 0 & 0 & 0 \\ 0 & 1 & 0 & 0 \\ 0 & 0 & 1 & 0 \\ 0 & 0 & 0 & 1 \end{pmatrix} \qquad (5)$$

leaves $M_k^{(0)}$ vector invariant, while, for example, the operator

$$\chi_2^k = \begin{pmatrix} 0 & 0 & 1 & 0 \\ 0 & 1 & 0 & 0 \\ 0 & 0 & 0 & 1 \\ 1 & 0 & 0 & 0 \end{pmatrix} \qquad (6)$$

acts as

$$\chi_2^k \begin{pmatrix} \Psi_{1k} \\ \Psi_{2k} \\ \Psi_{3k} \\ \Psi_{4k} \end{pmatrix} = \begin{pmatrix} \Psi_{3k} \\ \Psi_{2k} \\ \Psi_{4k} \\ \Psi_{1k} \end{pmatrix} \qquad (7)$$

giving a rotation of the above fundamental representation. It is worth stressing that $\det\|\chi_2^k\|=1$, so matrix (6) represents a rotation. On the other hand, a matrix operator of the form

$$\bar{\chi}_1^k = \begin{pmatrix} 0 & 0 & 0 & 1 \\ 0 & 1 & 0 & 0 \\ 0 & 0 & 1 & 0 \\ 1 & 0 & 0 & 0 \end{pmatrix} \tag{8}$$

generates the inversion between the bonds $\Psi_{1k}$ and $\Psi_{4k}$ with respect to the fundamental representation, that is

$$\bar{\chi}_1^k \begin{pmatrix} \Psi_{1k} \\ \Psi_{2k} \\ \Psi_{3k} \\ \Psi_{4k} \end{pmatrix} = \begin{pmatrix} \Psi_{4k} \\ \Psi_{2k} \\ \Psi_{3k} \\ \Psi_{1k} \end{pmatrix} \tag{9}$$

and $\det \left\| \bar{\chi}_1^k \right\| = -1$, which means that the application is an inversion. In general, the tetrahedral components of the chain (3) can be rotated or inverted in their bonds with respect to their fundamental representations and the matrices $\chi_q^k$ and $\bar{\chi}_q^k$ can be derived from Fischer projections [8]. We can give the general formula

$$M_n = \sum_{k=1}^{p} \bar{M}_k + \sum_{k=p+1}^{n} M_k \tag{10}$$

where $\bar{M}_k$ and $M_k$ are the k-tetrahedrons on which the matrix operators $\bar{\chi}_q^k$ and $\chi_q^k$ act respectively; k is the center index running from 1 to *n*; q is the operator index running from 1 to 12 [8]. The set of operators $\bar{\chi}_q^k$ and $\chi_q^k$ are elements of the algebra of the orthogonal group *O*(4). For any k-tetrahedron in the basic configuration (4), we can have the 24 representations given by

$$M_k = \chi_q^k M_k^{(0)}, \qquad \bar{M}_k = \bar{\chi}_q^k M_k^{(0)} \tag{11}$$

In this sense, Eq. (10) is nothing else but Eq. (3) where the action of transformations $\chi_q^k$ and $\bar{\chi}_q^k$ is specified. A given tetrahedral chain can be classified as

$$M_n = \sum_{k=1}^{n} M_k, \qquad p = 0 \qquad (12)$$

which is an achiral molecole on which only rotations $\chi_q^k$ acts on $M_k^{(0)}$;

$$M_n = \sum_{k=1}^{p} \overline{M}_k + \sum_{k=p+1}^{n} M_k, \qquad 0 < p < n \qquad (13)$$

is a diastereoisomer since $[n-(p+1)]$ tetrahedrons result superimposable after rotations, while $p$-ones are not superimposable, having, each of them, undergone an inversion of two of their bonds. In this case, the rotations $\chi_q^k$ and the inversions $\bar{\chi}_q^k$ act on the chain $M_n$.

Finally,

$$M_n = \sum_{k=1}^{n} \overline{M}_k, \qquad n = p \qquad (14)$$

is an enantiomer since the inversion operators $\bar{\chi}_q^k$ act on every tetrahedron.

## 3. The chirality index $\chi$ for a tetrahedral molecule

The above considerations directly lead to the definition of a global index capable of assigning the chirality of a given simply connected tetrahedral chain. Central chirality is then assigned by the number $\chi$ given by the couple $n, p$ that is

$$\chi \equiv \{n, p\} \qquad (15)$$

where $\chi$ is the chirality index, $n$ the number of stereogenic centers and $p$ the number of inversions (at most one for each tetrahedron), which $p$ of the $n$ tetrahedrons have undergone. The simply connected tetrahedral chains are immediately classified as $\chi \equiv \{n, 0\}$ *achiral molecules*; $\chi \equiv \{n, p < n\}$ *diastereoisomers*; and finally $\chi \equiv \{n, n\}$ *enantiomers*.

With these results in mind, we seek for an Aufbau procedure able to furnish indications on the global central molecular chirality of a compound. When adding up a new chiral tetrahedron to a given chain, the variation of $\chi$ (specifically of $p$) tell us whether the chiral character of the chain is changed or not.

## 4. An Aufbau approach for a tetrahedral chain

As we have seen, the chirality index $\chi$ allows an immediate chiral characterization of a given tetrahedral chain. The building-up process is then straightforward. Let us take into account a molecule, which is well-defined in its chiral feature, in the sense that, considering also its mirror image, it is clear to assess if the molecule is an enantiomer, a diastereoisomer or an achiral molecule. Then, let us add to this structure and its mirror image a further chiral center. In general, the resulting structure will be

$$\chi \equiv \{n+1, p+\Delta p\} \qquad (16)$$

where $\Delta p = 0,1$. The chiral properties of the new molecule are assigned by the $\Delta p$ value according to the following possibilities.

If $\Delta p = 0$, we can have

$$\chi_s \equiv \{n,0\} \implies \chi_f \equiv \{n+1,0\} \qquad (17)$$

in this case, the starting compound is an achiral molecule as well as the final one.

Again, for $\Delta p = 0$, we can have

$$\chi_s \equiv \{n, p\} \implies \chi_f \equiv \{n+1, p\} \qquad (18)$$

in this case, the starting molecule is a diastereoisomer, being $n > p$, as well as the final structure.

Finally, if

$$\chi_s \equiv \{n, n\} \Rightarrow \chi_f \equiv \{n+1, n\} \quad (19)$$

the starting molecule is an enantiomer, while the final one is a diastereoisomer.

If $\Delta p = 1$, the situations can be

$$\chi_s \equiv \{n, 0\} \Rightarrow \chi_f \equiv \{n+1, 1\} \quad (20)$$

from an achiral molecule, a diastereoisomer is obtained;

$$\chi_s \equiv \{n, p\} \Rightarrow \chi_f \equiv \{n+1, p+1\} \quad (21)$$

from a diastereoisomer, another diastereoisomer is obtained;

$$\chi_s \equiv \{n, n\} \Rightarrow \chi_f \equiv \{n+1, n+1\} \quad (22)$$

from an enantiomer, we get another enantiomer.

Eqs. (17)-(22) take into account all the possibilities, which can be easily iterated adding up any number of chiral centers to a given chain. In the general case, the *Aufbau rule* is

$$\chi \equiv \{n+n', p+p'\}; \forall n' \geq 1, \quad p' = \sum_{j=1}^{n'} \Delta p_j, \; \Delta p_j = 0, 1 \quad (23)$$

However, we have to consider that the rule works only for simply connected tetrahedral chains (even in the presence of atoms which act as spacers between the chiral centers), where chiral features are well-established with respect to the mirror image.

## 5. Discussion and Conclusions

In this paper, we developed a straightforward Aufbau approach in order to characterize the global central chirality of simply connected tetrahedral chains. We use a chirality index $\chi \equiv \{n, p\}$, where $n$ is the number of stereogenic centers and $p$ the number of inversions (at most one for each center). Adding up a chiral center to the structure gives rise to a new molecule, where $\chi \equiv \{n+1, p+\Delta p\}$. The fact that, in the addition, the variation of $p$ can be $\Delta p = 0,1$, assigns the chiral feature of the new compound. The process can be iterated to any number of chiral centers (always for simply connected tetrahedral chains).

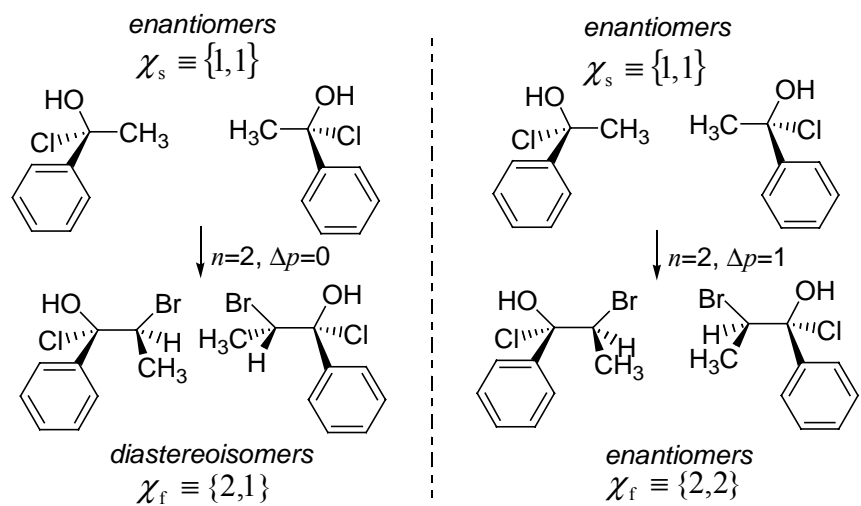

Fig. 2. Aufbau process consisting in adding up a chiral center to a given chiral tetrahedron.

An example of the building-up process is reported in Figure 2, while in Figure 3 is reported a degenerate case [9], where the two chiral centers are identical, introducing a further degree of symmetry to the final structure. In this case, the situation $\chi_f \equiv \{2,2\}$ for $\Delta p = 1$ is equivalent to $\chi_f \equiv \{2,0\}$, since the two molecules are superimposable, hence the structure is achiral. Last consideration indicates that such an Aufbau approach is working only if the chiral centers are different. In this sense, the procedure is an appropiate descriptor of chiral structures.

In conclusion, we have developed a method for the description of the central tetrahedral chirality by a chirality index, which not only assigns the global chirality of a given structure, but indicates also a way to predict the same property for new compounds, which can be built up.

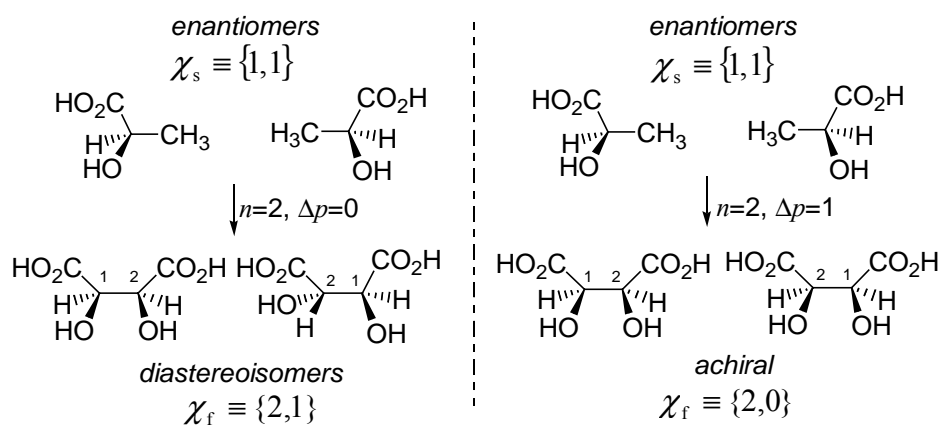

Fig. 3. A degenerate example of Aufbau process consisting in adding up a chiral center, having identical substituents of the starting chiral tetrahedron.